\documentclass[]{elsart}

\usepackage{times}
\usepackage{epsf}
\usepackage[numbers,sort&compress]{natbib}
\usepackage{amsmath}
\usepackage{amsfonts}
\usepackage{amssymb}
\usepackage{bm}
\usepackage{bbm}
\usepackage{graphicx}
\usepackage{epsfig}
\usepackage{latexsym}
\usepackage{nicefrac}

\def\corresponds{{\lower.2ex\hbox{=}}{\rm\kern-.75em^\triangle}}
\def\succsim{\succ\kern-.9em_\sim\kern.3em}
\def\precsim{\prec\kern-1em_\sim\kern.3em}
\def\slantfrac#1#2{\kern1em^{#1}\kern-.3em/\kern-.1em_{#2}}
\def\lfrac#1#2{{}^{#1\!}\kern-.0em/_{#2}}

\def\buildrel#1\under#2{\mathrel{\mathop{\kern0pt #2}\limits_{#1}}}

\def\tr{{\rm Tr}\,}

\newcommand{\beq}{\begin{equation}}
\newcommand{\eeq}{\end{equation}}
\newcommand{\bea}{\begin{eqnarray}}
\newcommand{\eea}{\end{eqnarray}}
\newcommand{\hf} {\frac{1}{2}}
\newcommand{\nonu}{\nonumber\\}

\newcommand{\sla}[1]{\mbox{$#1\!\!\!/$}}
\newcommand{\uvarphi}{{\underline \varphi}}

\def\ord#1{{\mathcal O}(#1)}

\def\mr#1{{\mathrm{#1}}}
\def\cF{{\mathcal F}}
\def\cL{{\mathcal L}}
\def\nn{\nonumber\\}

\begin{document}

\begin{frontmatter}
\title{Renormalization--Group Analysis\\
of Layered Sine--Gordon Type Models}

\author{I. N\'{a}ndori$^{1,2}$, S. Nagy$^{3}$, K. Sailer$^{3}$ and
U. D. Jentschura$^{2}$}

\address{
$^1$Institute of Nuclear Research of the Hungarian Academy of Sciences,\\
H-4001 Debrecen, P.O.Box 51, Hungary \\
$^2$Max--Planck--Institut f\"{u}r Kernphysik, Saupfercheckweg 1, 
69117 Heidelberg, Germany \\
$^3$Department of Theoretical Physics, University of Debrecen,
Debrecen, Hungary}
    
\date{\today}

\begin{abstract}
We analyze the phase structure and the renormalization group (RG) 
flow of the generalized sine-Gordon models with 
nonvanishing mass terms, using the Wegner-Houghton RG method in the 
local potential approximation. Particular emphasis 
is laid upon the layered sine-Gordon (LSG) model, which 
is the bosonized version of the multi-flavour Schwinger 
model and approaches the sum of two ``normal'',
massless sine-Gordon (SG) models in the limit of a vanishing 
interlayer coupling $J$. Another model of interest is 
the massive sine-Gordon (MSG) model.
The leading-order approximation to the 
UV (ultra-violet) RG flow predicts two phases
for the LSG as well as for the MSG,
just as it would be expected for the SG model, where the 
two phases are known to be separated by the Coleman fixed point.
The presence of finite mass terms (for the LSG and the MSG) 
leads to corrections
to the UV RG flow, which are naturally 
identified as the ``mass corrections''. The leading-order
mass corrections are shown to have the following consequences:
(i) for the MSG model, only one phase persists,
and (ii) for the LSG model, the transition temperature 
is modified. Within the mass-corrected UV scaling 
laws, the limit of $J \to 0$ is thus nonuniform with respect 
to the phase structure of the model. The modified
phase structure of general massive sine-Gordon models 
is connected with the breaking of 
symmetries in the internal space spanned by the field 
variables. For the LSG, the second-order subleading mass corrections 
suggest that there exists a cross-over regime before the 
IR scaling sets in, and the nonlinear terms show explicitly 
that higher-order Fourier modes 
appear in the periodic blocked potential.
\end{abstract}


\end{frontmatter}

\newpage

\section{Introduction}
\label{intro} 

At the heart of every quantum field theory, there is the need for 
renormalization. In the framework of the well-known perturbative 
renormalization procedure (see e.g. \cite{LGZJ1980,GuZJ1996}), the 
potentials---or interaction Lagrangians---are decomposed in a 
Taylor series in the fields; this Taylor series generates the vertices of 
the theory. If the expansion contains only a finite number 
of terms (this is the ``normal'' case), then each interaction vertex 
can be treated independently. However, certain theories exist which cannot 
be considered in this traditional way. In some theories, symmetries 
of the Lagrangian impose the requirement of taking  infinitely 
many interaction vertices into account; any truncation of these 
infinite series would lead to an unacceptable violation of 
essential symmetries of the model. The subject of this article is
to consider theories which fall into the latter category.

Specifically, we here consider generalizations of 
the well-known sine-Gordon (SG) scalar field theory
with mass terms. The ``pure,'' massless SG model 
is periodic in the internal space spanned by the field 
variable. One of the central subjects of investigation 
is the layered sine-Gordon (LSG) 
model~\cite{Fi1979,HeHoIs1995}, where the 
periodicity is broken by a coupling term between 
two layers each of which is described by a scalar field. 
All generalizations of the SG model discussed
here belong to a wider class of massive sine-Gordon type models for two
coupled Lorentz-scalar fields, which form an $O(2)$ ``flavour'' doublet,
i.e.~which are invariant under a global rotation in the 
internal space of the field variables, though not 
necessarily periodic. All Lagrangians investigated 
here contain self-interaction terms which are periodic in the 
field variables, but this periodicity is broken by the mass terms.

Regarding the phase structure, 
it is known that the massless sine-Gordon (SG) model for 
scalar, flavour singlet together with the two-dimensional 
XY model and Coulomb gas belong to the same universality class. For the 
two-dimensional Coulomb gas, the absence of long-range order, the 
existence of the Coleman fixed point and the presence of a topological 
(Kosterlitz--Thouless) phase transition have been proven rigorously 
in Refs.~\cite{MeWa1966,KoTh1973,Ko1974,JoKaKiNe1977,GeWe2000,sg2}.
It was shown that the dimensionful effective potential becomes a 
field-independent constant in both phases of the SG model~\cite{sg2}.

The joint feature of the massless and massive SG models is the 
presence of a self-interaction potential which is periodic in the 
various directions of the internal space. This makes it necessary to  
treat these models in a manner which avoids the Taylor-expansion of 
the periodic part of the potential. Hence, the 
renormalization~\cite{Wi1971,ZJRG,EyMoNiZJ1996,janosRG} of these 
models cannot be 
considered in the framework of the usual perturbative 
expansion~\cite{LGZJ1980,GuZJ1996}. The massive 
SG models open a platform to 
investigate the effect of a broken periodicity in the internal space. 
For the flavour singlet field, periodicity is broken 
entirely by a mass term, and the ground state is characterized by a 
vanishing field configuration~\cite{sanyi}.  

For the flavour doublet, one possible way to realize
a partial breaking of periodicity is given by a 
single nonvanishing mass eigenvalue. Alternatively, 
two eigenvalues of the ``mass matrix'' that enters the 
Lagrangian may be nonzero. We here
investigate the effect of entire and partial breaking of 
periodicity in the internal space on the ultraviolet (UV) scaling 
laws and on the existence of the Coleman fixed point. We shall 
restrict ourselves to various approximations  of the RG flow 
equation for the blocked potential.

The LSG model, because of its layered structure,
has a connection to solid-state physics.
In particular, it has been used to describe the 
vortex properties of high transition-temperature superconductors 
(HTSC)~\cite{pierson,pierson_map,pierson_lsg,pierson_rg,lsg_rg}.
The real-space renormalization group (RG) analysis of the LSG model,
invariably based on the dilute vortex gas approximation, 
has been successfully applied 
for the explanation of electric transport properties of HTSC 
materials \cite{pierson,pierson_lsg,lsg_rg,csmag}. New 
experimental data are in disagreement with theoretical predictions, 
and this aspect may require a more refined 
analysis as compared to the dilute gas approximation~\cite{kalman,csmag}. 

There exist connections of the generalized sine-Gordon models to 
fundamental questions of field theory.
For instance, a special case of the massive SG-type models 
is just the bosonized version of the massive Schwinger model,
which in turn is an exactly solvable 
two-dimensional toy-model of strong confining 
forces~\cite{HeHoIs1995,Fi1979}. 
The flavour singlet field can then be considered a meson field 
with vanishing flavour charge (``baryon number''), while 
the flavour doublet field models ``baryons'' with ``baryon charge''
$\pm\hf$.  Here, we restrict ourselves to the investigation of the 
vacuum sector with zero total flavour charge (``baryon charge'') 
\cite{DeSc1997,SmVe1996}. 
Of fundamental importance is the following question:
Are there any operators, irrelevant 
in the bare theory, which become relevant for the infrared (IR) physics?
Our investigations hint at some interesting 
phenomena which are connected with cross-over regions
in which UV-irrelevant couplings may turn into IR-relevant operators,
after passing through intermediate scales.
The IR-relevant ``confining forces'' would correspond to 
the interactions among the ``hadrons''
in our language. In the case of QCD, 
the much more serious problem of the determination of the 
operators relevant for confinement (i.e., for building up the hadrons) 
may, in principle, carry some similarities to the model 
problems studied here.

Our paper is organized as follows. In Sec.~\ref{def}, we give a short 
overview of all classes of massive generalized sine-Gordon 
models, of the flavour-doublet type, which are relevant 
for the current investigation, including the LSG and 
the MSG models. Section~\ref{nonpert} includes the basic relations used for the 
Wegner-Houghton (WH) RG method~\cite{wh} 
in the local potential approximation. In Sec.~\ref{rgflow}, we start with 
the outline of various approximations to the 
WH--RG used in the present paper. 
The UV scaling laws for the massless and massive models are found 
analytically in subsections~\ref{uvml} and \ref{nonlinrg}, 
respectively. In subsection \ref{nonlinrg}, the existence of the 
Coleman fixed point in massive SG models is also discussed on the 
basis of the UV scaling laws for various special cases, with entire 
and partial breaking of periodicity, for flavour-doublet and 
flavour-singlet fields. 
In Sec.~\ref{nonlinflowlsg}, the UV scaling laws are 
enhanced by keeping the subleading 
nonlinear terms in the mass-corrected RG flow equation for 
the blocked potential. In this approximation, the numerical determination 
of the RG flow is presented for the LSG model, and the existence of a 
cross-over region from the UV to the IR scaling regimes is
demonstrated to persist after the inclusion of the 
subleading terms. Finally, the main results are 
summarized in Sec.~\ref{conclu}. 

\section{Two-flavour Massive sine-Gordon Model}
\label{def}

In this article, we investigate a class of Euclidean scalar models for  
the flavour $O(2)$-doublet 
\begin{equation}
\uvarphi= \begin{pmatrix} \varphi_1\cr
\varphi_2 \end{pmatrix}
\end{equation}
in $d=2$ spatial dimensions. The bare Lagrangians are assumed to 
have the following properties:
\begin{enumerate}
\item The Lagrangians has the discrete symmetry 
$\uvarphi \to -\uvarphi$ (G-parity).
\item The flavour symmetry $\varphi_1 \longleftrightarrow \varphi_2$
leaves the Lagrangian invariant.
\item The Lagrangian contains an interaction term $U(\varphi_1,\varphi_2)$,
periodic in the internal space spanned by the field 
variables,
\begin{equation}
\label{property2}
U(\varphi_1,\varphi_2)= 
U\left(\varphi_1+ \frac{2\pi}{b_1},\varphi_2+\frac{2\pi}{b_2}\right)\,,
\end{equation}
with $b_i = {\rm const.}$ (for $i=1,2$). 
As shown below, we may even assume $b_1 = b_2$
without loss of generality.
\item The Lagrangian contains a mass term 
$\hf \uvarphi^\mr{T} {\underline{\underline M}}^2\uvarphi$,
where the symmetric, positive semidefinite 
mass matrix $M^2_{ij}$ $(i,j=1,2)$ has the structure
\begin{equation}
\label{property3}
{\underline{\underline M}}^2= 
\left( \begin{array}{cc} M_1^2 &\hspace*{0.5cm} -J\\
-J &\hspace*{0.5cm} M_2^2 \end{array} \right) \,, \qquad
\det {\underline{\underline M}}^2 \geq 0\,,
\end{equation}
with $M_1^2,~M_2^2,~J\ge 0$. Flavour symmetry 
imposes the further constraint $M_1 = M_2$, but 
initially we will prefer to 
keep an arbitrary $M_1$ and $M_2$ in the 
formulas, for illustrative purposes.
\end{enumerate} 
We will call a general 
Lagrangian having the above properties a general
\begin{center}
{\em two-flavour massive sine-Gordon model} (2FMSG).
\end{center}
Various specializations will be discussed below.
Invoking the completeness of a Fourier decomposition, we see immediately
that the general structure of the bare action of 
a 2FMSG model is
\begin{eqnarray}\label{clb}
\cL_b &=& 
{1\over 2} (\partial {\uvarphi}^\mr{T}) (\partial {\uvarphi})
+ {1\over 2}{ \uvarphi}^\mr{T} {\underline {\underline M}}^2 { \uvarphi}
\nonumber\\ 
&+& \sum_{n,m=0}^{\infty} 
\left[f_{nm} \cos(nb_1 \,\varphi_1)\cos(mb_2 \,\varphi_2) +
g_{nm} \sin(nb_1 \,\varphi_1)\sin(mb_2 \,\varphi_2)
\right].
\end{eqnarray}
Here, all couplings $f_{nm}$ and $g_{nm}$ are dimensionful (the dimensionless
case will be discussed below).

Some of the Lagrangians we will consider actually depend 
on one flavour only. For these, the flavour symmetry
requirement~(2) is not applicable.

An orthogonal transformation
\begin{equation}\label{rotation}
{\underline {\underline{\mathcal O}}}=
\begin{pmatrix} \cos \gamma & \sin \gamma\cr
               -\sin \gamma & \cos \gamma 
\end{pmatrix}
\end{equation}
of the flavour-doublet, 
$\uvarphi \to{\underline {\underline {\mathcal O}}}\, \uvarphi$,
transforms the model into a similar one with transformed period
lengths in the internal space,
\begin{equation}
\begin{pmatrix} \beta^{-1}_1 \cr \beta^{-1}_2 \end{pmatrix}
= 
\begin{pmatrix} \cos \gamma & \sin \gamma\cr
                 -\sin \gamma & \cos \gamma 
\end{pmatrix}
\begin{pmatrix} b^{-1}_1 \cr b^{-1}_2 \end{pmatrix}\,.
\end{equation}
There exists a particular orthogonal 
transformation, the rotation by the angle 
\begin{equation}
\gamma_{12} = \arctan\left(\frac{b_1-b_2}{b_1+b_2}\right)\,,
\end{equation}
which transforms the periodic structure to the 
case of equal periods $\beta_1=\beta_2=\beta$,
\begin{eqnarray}\label{cl}
\cL &=& {1\over 2} \, (\partial {\uvarphi}^\mr{T}) (\partial {\uvarphi})
+ {1\over 2} \, { \uvarphi}^\mr{T} {\underline {\underline M}}^2 
{ \uvarphi}
\nonumber\\ 
&+& \sum_{n,m=0}^{\infty} 
\left[u_{nm} \cos(n\beta \, \varphi_1)\cos(m\beta \, \varphi_2) +
v_{nm} \sin(n\beta \, \varphi_1)\sin(m\beta \, \varphi_2)
\right].
\end{eqnarray}
For the sake of simplicity, we did not change the notations for the
transformed (rotated) field and mass matrix. However, the couplings are
now denoted as $u_{nm}$ and $v_{nm}$. The scaling laws do not differ 
qualitatively for the model $\cL_b$ [see Eq.~(\ref{clb})]
with different periods in the 
different directions of the internal space on the 
one hand, and for $\cL$ [see Eq.~(\ref{cl})] with an
identical period $\beta$ in both directions of the internal space
on the other hand. The global $O(2)$ rotation in
Eq.~(\ref{rotation}), which connects these 
bare theories, does not mix the field fluctuations with different 
momenta, so that the same global rotation connects the blocked 
theories at any given scale. Without loss of generality,
we may therefore 
restrict our considerations below to the models with identical periods 
in both directions of the internal space.

For the model given by the Lagrangian $\cL$ of Eq.~(\ref{cl}),
the positive semidefinite mass matrix has the eigenvalues,
\begin{equation}
\label{defD}
M_\pm^2 = \frac{M_1^2+M_2^2}{2} \pm \biggl[
\biggl( \frac{M_1^2-M_2^2}{2} \biggr)^2 + J^2 \biggr]^{\hf} 
= T\pm D \ge 0.
\end{equation}
we may now distinguish the following cases:
\begin{itemize}
\item case (i): two vanishing eigenvalues $M_\pm^2 =0$, 
\item case (ii): $M_-^2=0$, but $M_+^2=2M^2=2J>0$, and
\item case (iii): two nonvanishing eigenvalues $M^2_\pm \neq 0$. 
\end{itemize}
Case (i) occurs for $M_1^2=M_2^2=J=0$ and represents
the {\em massless two-flavour SG model (ML2FSG)}.
Case (ii) is relevant for $M_1^2=M_2^2=J\not=0$,
and case (iii) occurs for  $M_1^2\,M_2^2>J^2$.
In case (i), the periodicity in the 
internal space is fully respected by the entire Lagrangian
[not only by its periodic part, see Eq.~(\ref{cl})].
by contrast, cases (ii) and (iii) correspond to  explicit breaking of 
periodicity either partially or entirely, respectively. This is 
because one could have diagonalized the mass matrix in the latter 
case by an appropriate $O(2)$ rotation, in which case
one would have arrived at a Lagrangian 
of the form of Eq.~(\ref{clb}) for which the mass term would break 
periodicity either in a single direction, or both (orthogonal) 
directions in the internal space. 

In the bare potential, we will assume a simple structure 
for the periodic part [which is the part which containing 
the $u_{nm}$'s and $v_{nm}$'s in Eq.~(\ref{cl})].
Indeed, we will restrict ourselves to only 
one nonvanishing Fourier mode with indices $(n,m)=(1,0)$
in the periodic part of the bare potential in the 
Lagrangian $\cL$. By choosing a particular angular phase
for the field variable, we can restrict the discussion to the 
$u$-mode and ignore the $v$-mode. Note that 
because of flavour symmetry, we could have
chosen $(n,m)=(0,1)$ as well, $u_{10} = u_{01}$.
Applying this special structure,
we recover various models of physical interest:
\begin{enumerate}
\item Respecting global flavour symmetry 
$\varphi_1 \longleftrightarrow \varphi_2$,
the choice $M_1^2=M_2^2$, together with the 
restriction to only one Fourier mode,
results in the {\em symmetric 2FMSG model} (S2FMSG).
The Lagrangian reads
\begin{eqnarray}\label{s2fmsg}
{\mathcal L}_{\rm S2FMSG} &=& 
{1\over 2} (\partial \varphi_1)^2 + {1\over 2} (\partial \varphi_2)^2
- {J} \varphi_1\varphi_2 \nonumber\\[2ex]
&+& {1\over 2} {M}^2 (\varphi_1^2 + \varphi_2^2)  
+ {u} \left[\cos(\beta\varphi_1) + \cos(\beta\varphi_2)\right]\,.
\end{eqnarray}
Here, the notations ${M}^2 \equiv {M}_1^2 = {M}_2^2$ and 
$u \equiv u_{01} = u_{10}$ are introduced. The mass eigenvalues are
$M_\pm^2={M}^2 \pm J\ge 0$ (because we assume a positive 
semidefinite mass matrix). 
For $M_\pm^2={M}^2 \pm J > 0$,
the S2FMSG model belongs to case~(iii).
\item We now specialize the S2FMSG model to the 
case $J=M_1^2=M_2^2$ with mass eigenvalues $M_+^2=2J>0$ and $M_-^2=0$.
This yields the layered sine-Gordon model (LSG),
which belongs to the case (ii) in the above 
classification, and the 
Lagrangian reads
\begin{equation}\label{lsg}
{\mathcal L}_{\rm LSG} =
{1\over 2} \, (\partial \varphi_1)^2 + 
{1\over 2} \, (\partial \varphi_2)^2
+ {1\over 2} \, J(\varphi_1 -\varphi_2)^2 
+ {u} \left[\cos(\beta\, \varphi_1) + \cos(\beta\,\varphi_2)\right].
\end{equation}
The LSG model has been used to describe the vortex properties of 
high-transition temperature superconductors 
(HTSC)~\cite{pierson,pierson_map,pierson_lsg,pierson_rg,csmag,kalman,lsg_rg}.
Typical HTSC materials have 
a layered microscopic structure. In the framework 
of a (layered, modified) Ginzburg-Landau theory of superconductivity, 
the vortex dynamics of strongly anisotropic HTSC materials can be 
described reasonably well by the layered XY or layered vortex (Coulomb) 
gas models, which in turn can be mapped onto the LSG model. The 
adjacent layers are treated on an equal footing, and the mass term
$+ \hf \, J (\varphi_1-\varphi_2)^2$ describes the weak interaction of 
the neighbouring layers. The parameter $\beta$ is related to the 
inverse-temperature of the layered system~\cite{pierson_lsg}.

The particular choice of $\beta =2\sqrt{\pi}$ for the LSG
represents the bosonized version of the two-flavour massive Schwinger 
model (c.f. Appendix \ref{2fmsm}).
\item Equation (\ref{s2fmsg}),
for $M = J = 0$, represents the {\em massless two-flavour 
sine-Gordon model} (ML2FSG). Periodicity in the internal space is fully 
respected.
\item The Lagrangian in Eq.~(\ref{s2fmsg}), with 
$J=0$ and $M_1^2=M^2\not=0$, 
$M_2^2=0$ gives the Lagrangian 
${\mathcal L}_{\rm MSG}$ of the 
(one-flavour) {\em massive sine-Gordon model} (MSG),
\begin{eqnarray}\label{msg}
{\mathcal L}_{\rm MSG} = {1\over 2} (\partial \varphi)^2 
+   {1\over 2} {M}^2 \, \varphi^2 
+ {u} \cos(\beta\varphi).
\end{eqnarray}
For the other massless scalar field, a massless theory results. 
It is well-known, 
that the MSG model for $\beta=2\sqrt{\pi}$ is the bosonized 
(one-flavour) massive Schwinger model~\cite{Co1973,Co1975,Co1976}.
In the language of Appendix~\ref{2fmsm}, the one-flavour
model would correspond to Eq.~(\ref{2dqed}) with the sum
over $i$ restricted to a single term.
\end{enumerate}

\section{Wegner-Houghton's RG Approach in Local Potential Approximation}
\label{nonpert}

The critical behaviour and phase structure of the LSG-type models have 
been investigated by several perturbative (linearized) methods (see 
e.g.~\cite{Co1976,Fi1979,pierson,pierson_map,pierson_lsg,pierson_rg}),
providing scaling laws, which {\em a priori} are valid in UV. 
Here, our purpose is to go beyond 
the linearized results and to obtain scaling laws for 
specializations of the 2FMSG model,
the validity of which is extended from the UV region towards the scale 
of the mass eigenvalues. 

We apply a differential RG in 
momentum space with a sharp cut-off $k$, the so-called Wegner-Houghton
RG approach to the general 2FMSG model. In principle,
this method (in its nonlinearized, full version) 
enables one to determine the blocked action down to the IR limit $k\to 0$.
The blocked action $S_k [\uvarphi]$ at the momentum scale $k$ is 
obtained from the bare action $S_\Lambda[\uvarphi] $ at the 
UV cut-off scale $\Lambda$ by integrating out the  high-frequency modes 
of the field fluctuations above the moving cut-off $k$. Performing 
the elimination of the high-frequency modes successively, in momentum 
shells $[k-\Delta k, k]$  of infinitesimal thickness $\Delta k\to 0$, 
the following integro-differential equation is obtained,
\begin{equation}\label{bl} 
k \, \partial_k S_k [\uvarphi] = -\lim_{\Delta k \to 0} \frac{1}{2\Delta k}
\tr' \ln S_k^{ij} [\uvarphi]\,.
\end{equation}
The WH equation is a so-called exact RG flow equation for the blocked action.
The trace $\tr'$ on the right hand side has to be taken over the  modes
with momenta in the  momentum shell $[k-\Delta k, k]$. 
We shall assume bare couplings for which the second functional 
derivative matrix 
\begin{equation}
S_k^{ij}[\uvarphi] = \frac{\delta^2 S_k[\uvarphi]}
  {\delta \varphi_i\delta\varphi_j}
\end{equation}
remains positive definite in the UV scaling region, so that the flow 
equation (\ref{bl}) does not lose its validity due to the so-called 
spinodal instability. Blocking generally affects physics which is reflected 
in the scale-dependence of the couplings of the blocked action.

The WH-RG equation (\ref{bl}) has to be projected onto a particular 
functional subspace, in order to reduce the search for a functional 
(the blocked action) to the determination of the flow of 
{\em coupling parameters} that multiply functions of the field variables
(see also Appendix~\ref{WHapp}).
Here, we assume that the 
blocked action contains only local interactions and restrict ourselves to 
the lowest order of the gradient expansion, the so-called local potential
approximation (LPA)~\cite{janosRG,ZJRG}, according to 
which the fields remain constant over all space. We assume that the Lagrangian
of the blocked theory is of the same  form as that of the bare theory
$\cL$ of Eq. (\ref{cl}), but with scale-dependent parameters.

We introduce the dimensionless blocked potential
$\tilde V_k (\varphi_1, \varphi_2) = k^{-2}\, V_k(\varphi_1, \varphi_2)$,
dimensionless mass parameters 
$\tilde M^{ij}_k = k^{-2}\, M^{ij}_k$
and couplings ${\tilde u}_{ij} = k^{-2}\, u_{ij}$.
All dimensionless quantities will be denoted by a tilde superscript
in the following. 
We recall that in $d = 2$ dimensions, the fields have carry
no physical dimension,
so that $\underline\varphi = \underline{\tilde\varphi}$. 

As already emphasized [see Eq.~(\ref{cl})], 
throughout this article we assume that the 
dimensionless potential ${\tilde V}_k$ is the 
sum of the dimensionless mass term [proportional
to $\uvarphi^\mr{T} {\underline {\underline {\tilde M}}}^2 (k) \uvarphi$]
and of the dimensionless periodic potential 
${\tilde U}_k(\varphi_1, \varphi_2)$,
\begin{equation}\label{def1}
{\tilde V}_k (\varphi_1, \varphi_2)= \hf \,
\uvarphi^\mr{T} {\underline {\underline {\tilde M}}}^2 (k) \uvarphi 
+ {\tilde U}_k(\varphi_1, \varphi_2).
\end{equation}
In the language of Eq.~(\ref{bl}),
we obtain $S_k^{ij}= \delta^{ij}+ {\tilde V}_k^{ij}$,
and the following equation (again for $d=2$, see Ref.~\cite{lsg_rg}),
\begin{eqnarray}
\label{WHdim}
\lefteqn{
\left(2 + k \, \partial_k \right) {\tilde V}_k (\varphi_1, \varphi_2)} 
\nonumber\\
& =& - \alpha_2 \ln\left(
[1 +  {\tilde V}^{11}_k(\varphi_1, \varphi_2) ] 
[1 + {\tilde V}^{22}_k(\varphi_1, \varphi_2) ] 
-  [ {\tilde V}^{12}_k(\varphi_1, \varphi_2)]^2 \right) \,,
\end{eqnarray}
where the notation
\begin{equation}
\label{defderiv}
{\tilde V}^{ij}_k(\varphi_1, \varphi_2) \equiv 
\partial_{\varphi_i}\partial_{\varphi_j}
{\tilde V}_k(\varphi_1, \varphi_2)
\end{equation}
is used for the second derivatives
with respect to the fields in Eq.~(\ref{WHdim}). 
The numerical constant $\alpha_2= 1/(4\pi)$,
is a specialization of the general form
\begin{equation}
\label{alphad}
\alpha_d = \frac{\Omega_d}{2\,(2\pi)^d}
\end{equation}
to the case $d=2$. Here,
\begin{equation}
\Omega_d = \frac{2\,\pi^{d/2}}{\Gamma(d/2)}
\end{equation}
is the $d$-dimensional solid angle. 

We recall that in the LPA, the blocked potential  
${\tilde V}_k(\varphi_1, \varphi_2)$ is a function of the real 
variables (constant field configurations) $\varphi_i$, $(i=1,2)$. 
The scale-dependence is entirely encoded in the dimensionless coupling 
constants of the blocked potential. Inserting the ansatz (\ref{def1}) 
into the WH-RG equation (\ref{WHdim}), the right hand side turns out to 
be periodic, while the left hand side contains both 
periodic and non-periodic 
parts. The non-periodic part contains the 
mass term, and we obtain the trivial tree-level evolution
for the dimensionless mass parameters
${\tilde M}^2_{ij}(k)$,
\begin{eqnarray}
\label{treelevel}
{\tilde M}^2_{ij}(k) &=& 
{\tilde M}^2_{ij} (\Lambda) \left({k\over\Lambda}\right)^{-2}
\end{eqnarray}
and the RG flow equation
\begin{eqnarray}
\label{Uflow}
\lefteqn{
\left(2 + k \, \partial_k \right) {\tilde U}_k (\varphi_1, \varphi_2)} 
\nonumber\\
& =& - \alpha_2 \ln\left([1 +  {\tilde V}^{11}_k(\varphi_1, \varphi_2)] 
[1 +  {\tilde V}^{22}_k(\varphi_1, \varphi_2) ] 
- [{\tilde V}^{12}_k(\varphi_1, \varphi_2)]^2 \right) \,
\end{eqnarray}
for the  dimensionless periodic piece of the blocked potential.
Hence, the dimensionful mass parameters $ M^2_{ij} = k^2
{\tilde M}_{ij}^2(k)$ remain constant during the blocking. It is 
important to note that the RG flow equation (\ref{Uflow}) keeps the 
periodicity of the periodic piece $\tilde U_k$ of the blocked 
potential in both directions of the internal space with unaltered 
length of period $\beta$.

\section{RG Flow}
\label{rgflow}

%
%
\subsection{Orientation}

We wish to concentrate on the 
scaling laws in the UV region and their extension toward the scale of 
the largest eigenvalue of the mass matrix.
First, we determine the UV scaling laws for the corresponding
massless models. For this purpose, the RG-flow equation 
(\ref{Uflow}) is linearized in the full potential,
by expansion of the logarithm,
\begin{equation}\label{WHlin1}
\left(2 + k \, \partial_k \right) {\tilde U_k(\varphi_1,\varphi_2)} =
\,-  \alpha_2 \, \left({\tilde V^{11}_k} + {\tilde V^{22}_k}\right).
\end{equation}
The linearization is valid 
provided the inequalities $|{\tilde V^{ij}_k}|\ll 1$ hold.
This approximation is applicable in the UV, 
because the dimensionless ${\tilde V^{ij}}_k$
are obtained from the dimensionful as $V^{ij}_k$ by a 
multiplicative factor $k^{-2}$.
The solution of Eq. (\ref{WHlin1}) provides the correct scaling 
laws for massless models like the ML2FSG.
The mass terms enter Eq.~(\ref{WHlin1}) only via 
a $k$-dependent, but field-independent term on the 
right hand side and do not influence the 
RG flow of the coupling parameters ${\tilde u}_{nm}$
and ${\tilde v}_{nm}$ that enter the
periodic part of the potential.

Second, we determine the UV scaling laws for the massive models. 
We assume 
\begin{equation}
|{\tilde U^{11}_k}+ {\tilde U^{22}_k} +\ord{ (\tilde V^{ij}_k)^2 } |\ll
1 + {\tilde \mu}^2, \qquad
{\tilde \mu}^2={\mbox {~tr~}}{\tilde M}_{i,j}^2
+{\mbox {~det~}}{\tilde M}_{i,j}^2\,,
\end{equation}
and expand  the logarithm in the right hand side of Eq.~(\ref{Uflow}), 
\begin{eqnarray}
\label{logexpand}
\lefteqn{
\ln [ 1+ {\tilde \mu}^2 + {\tilde U^{11}_k}+ {\tilde U^{22}_k}
+\ord{ (\tilde V^{ij}_k)^2 } ] }
\nonumber\\
&\approx&
\ln \biggl( 1 + 
\frac{ {\tilde U^{11}_k}+ {\tilde U^{22}_k}
+ \ord{\tilde V^{ij}_k)^2}}{1+ \tilde \mu^2 }
  \biggr) + \ln \left( 1 + \tilde\mu^2 \right)
 \nonu
 &=& \cF_1(\tilde U_k)+\cF_2(\tilde U_k)+\ldots +  
\ln\left( 1 + \tilde\mu^2 \right)\,.
\end{eqnarray}
The terms $\cF_1(\tilde U_k)$ and $\cF_2(\tilde U_k)$ 
represent the linear and quadratic terms
in the second derivatives of the periodic potential, respectively,
obtained by expansion of the logarithm.
These terms are given explicitly in Eq.~(\ref{cF1cF2}) below.
Note that $\tilde \mu^2 \ge 0$ holds for a 
positive semidefinite mass matrix.
In view of the structure of the two-flavour WH-equation (\ref{Uflow}),
one can add and subtract, on the right-hand side, a field-independent,
but possibly $k$-dependent term without changing the 
RG evolution of the coupling constants. This term may be 
chosen as $\ln\left( 1 + \tilde\mu^2 \right)$, because of the 
trivial RG evolution of the mass terms in
Eq.~(\ref{treelevel}).

The mass-corrected RG flow equation
\begin{equation}\label{uflex}
(2 + k\,\partial_k ) \tilde U_k (\varphi_1,\varphi_2)  =
-\alpha_2 \lbrack {\cF}_1 (\tilde U_k) +  {\cF}_2 (\tilde U_k) + \ldots \rbrack
\end{equation}
is obtained. The mass corrections help in extending the range
of validity of the UV scaling laws of the general 2FMSG model
towards the scale $k\sim \ord{M_+}$. A better 
approximation can be achieved by using both the linear 
and the quadratic terms ${\cF}_1 (\tilde U_k)$ and 
${\cF}_2 (\tilde U_k)$ instead of the linear 
terms only. Because of the tree-level evolution 
(\ref{treelevel}), $\tilde \mu \to 0$ for $k \to\infty$,
and thus, the mass corrections vanish in the UV.
All of these approximation schemes are illustrated in the following.

\subsection{UV scaling laws for massless models}
\label{uvml}

As argued above, the UV scaling laws of the massive models in the 
extreme UV limit, $\Lambda \sim k\gg M_+$, are asymptotically 
equivalent to those of 
the corresponding massless models. The  UV scaling laws of the ML2FSG 
model are obtained by solving the linearized RG equation (\ref{WHlin1}),
which results in decoupled flow equations for the various Fourier 
amplitudes. Their solutions can be obtained analytically,
\begin{eqnarray}\label{linsol} 
\begin{pmatrix}
{\tilde u}_{nm}(k) \cr {\tilde v}_{nm}(k)\end{pmatrix}
& =&\left({k\over\Lambda}\right)^{-2 + \alpha_2\, \beta^2 (n^2 +m^2 )}
 \begin{pmatrix} {\tilde u}_{nm}(\Lambda) \cr
   {\tilde v}_{nm}(\Lambda) \end{pmatrix}.
\end{eqnarray}
Here, ${\tilde u}_{nm}(\Lambda)$ and ${\tilde v}_{nm}(\Lambda)$ are 
the initial values for the coupling constants at the UV cutoff $\Lambda$,
and we recall that $\alpha_2 = 1/(4\pi)$ has got nothing to do 
with a coupling constant [see Eq.~(\ref{alphad})]. 
We immediately see that the linearized RG flow predicts a Coleman-type 
fixed point for the ML2FSG model with a single Fourier mode 
($n = 0$, $m=1$) of the 
potential at the critical value $\beta^2_c = 8\pi$. A similar fixed point 
was found in the massless sine-Gordon model \cite{sg2,sg3}. For the 
ML2FSG model with infinitely many  Fourier modes of the periodic 
potential, all the Fourier amplitudes ${\tilde u}_{nm}(k)$  and 
${\tilde v}_{nm}(k)$ are UV irrelevant for $\beta^2 >\beta^2_c$, while  
for $\beta^2 <\beta^2_c $, at least one of the Fourier amplitudes 
becomes relevant. However, one should 
remember that on the basis of the linearized RG flow equation,
it is hardly possible to make any definite conclusion regarding the 
existence of a Coleman-type fixed point for massive sine-Gordon type 
models, since the linearized RG flow equation takes into account 
neither the effects of the finite mass eigenvalues, nor those of the 
nonlinear terms which couple the various Fourier amplitudes of the 
blocked potential. We therefore cannot use Eq.~(\ref{WHlin1}) 
or~(\ref{linsol}) for a description of the phase 
structure of the massive models, although the mass-corrected 
flow (\ref{uflex}) reproduces the massless flow (\ref{WHlin1}) 
in the ``extreme UV,''
which might be called the ``XUV region'' in some distant analogy to the 
corresponding short wavelengths of light.

\subsection{Mass-corrected UV scaling laws for massive models}
\label{nonlinrg}

In the case of general 2FMSG models, the mass parameters $\tilde J(k)$, 
$\tilde M_1^2 (k)$ and $\tilde M_2^2 (k)$ are always relevant in the IR
[see Eq.~(\ref{treelevel})]. 
This means that the argument of the logarithm in Eq.~(\ref{Uflow}) will 
always increase for decreasing scale $k$, regardless of the
choice of  the initial conditions for the coupling constants.
Consequently, the linearization (\ref{WHlin1}) necessarily 
loses its validity with decreasing scale $k$, irrespective of the 
value of $\beta$. This observation suggests that
one has to turn to  Eq. (\ref{uflex}), in order to extend 
the scaling laws towards the  scale $k \sim\ord{M_+}$. 
By contrast, for the ML2FSG model there are no mass terms, and the
linearization may remain valid down to  the IR limit (if $\beta^2 >
 \beta_c^2$).

The detailed evaluation of the terms in the right hand side of
Eq. (\ref{uflex}) gives
\begin{subequations}
\label{cF1cF2}
\begin{eqnarray}
\label{cF1cF2a}
{\cF}_1 (\tilde U_k) &=& 
r_1\, \tilde U^{11}_k + r_2\, \tilde U^{22}_k - 2r\, 
\tilde U^{12}_k ,\\
\label{cF1cF2b}
{\cF}_2 (\tilde U_k) &=&
-\hf\, r_1^2 [ \tilde U^{11}_k]^2
-\hf\, r_2^2 [ \tilde U^{22}_k]^2 
- ( \xi+2r^2 ) [ \tilde U^{12}_k]^2
- r^2\, \tilde U^{11}_k \tilde U^{22}_k
\nonumber\\
&& + 2 r_1 r \, \tilde U^{11}_k \tilde U^{12}_k
+ 2 r_2 r \, \tilde U^{22}_k \tilde U^{12}_k
\end{eqnarray}
with
\begin{align}
\label{cF1cF2c}
\xi = (1+\mu^2)^{-1}, & \qquad r= \xi \tilde M_{12}^2\,,
\nonumber\\
r_1 = \xi( 1+ \tilde M_{22}^2 ), & \qquad
r_2 = \xi ( 1+ \tilde M_{11}^2 )\,.
\end{align}
\end{subequations}

For the remainder of the derivation, we will 
restrict our attention to the linear term ${\cF}_1(\tilde U_k)$ on 
the right hand side of Eq.~(\ref{uflex}) and equate the coefficients 
of the corresponding Fourier modes on both sides of the equation.
We will assume a Lagrangian of the general structure
\begin{eqnarray}
\label{genstruc}
{\mathcal L} &=& 
{1\over 2} (\partial \varphi_1)^2 + {1\over 2} (\partial \varphi_2)^2
- {J} \varphi_1\varphi_2 \nonumber\\[2ex]
&+& {1\over 2} M_1^2 \varphi_1^2 + {1\over 2} M_2^2 \varphi_2^2
+ {u} \left[\cos(\beta\varphi_1) + \cos(\beta\varphi_2)\right]\,,
\end{eqnarray}
which is almost equivalent to the S2FMSG model as defined 
in Eq.~(\ref{s2fmsg}), but we keep two different 
masses $M_1$ and $M_2$, for illustrative purposes.
 
One finally arrives at the following 
set of equations for the scale-dependent Fourier 
amplitudes, 
\begin{eqnarray}
{\bf D}_k 
\begin{pmatrix}\tilde u_{nm} \cr \tilde v_{nm} \end{pmatrix}
&=& 
\alpha_2 \, \beta^2
\begin{pmatrix} {\bf A} &-{\bf B} \cr
-{\bf B} & {\bf A} \end{pmatrix} 
\begin{pmatrix}\tilde u_{nm} \cr \tilde v_{nm} \end{pmatrix} \,.
\label{nonlin_nm}
\end{eqnarray}
Here, the differential operator ${\bf D}_k \equiv 2+k \,\partial_k$,
and the coefficients are
\begin{equation}
{\bf A} = \frac{(1+\tilde M_1^2)\,m^2+
(1+\tilde M_2^2)\,n^2}{(1+\tilde M_1^2)
(1+\tilde M_2^2)-\tilde J^2},\quad
{\bf B} = \frac{2 n m \, \tilde J}{(1+\tilde M_1^2)
(1+\tilde M_2^2)-\tilde J^2}.
\end{equation}
We see that modes given by different pairs of integers $(n,m)$
decouple due to the linearization, but the corresponding cosine and 
sine modes mix. The set of Eqs.~(\ref{nonlin_nm}) decouple entirely when the 
functions
\begin{equation}
\tilde F_{\pm~nm} = \tilde u_{nm}\pm\tilde v_{nm}
\end{equation}
are introduced,
\begin{equation}
{\bf D}_k \tilde F_{\pm~nm} 
= \alpha_2\beta^2\left({\bf A}\mp{\bf B}\right)\tilde F_{\pm~nm}.
\end{equation}
The solution is easily found to be 
\begin{eqnarray}
\tilde F_{\pm~nm}(k)
 &=& \tilde F_{\pm~nm}(\Lambda)\left(\frac{k}{\Lambda}\right)^{-2}
\prod_{\lambda=\pm}
[R_\lambda (k)]^{\alpha_{nm} +\lambda(\beta_{nm}
\pm\gamma_{nm} )} 
\end{eqnarray}
with the variables
\begin{equation} 
R_\lambda (k)=\frac{k^2+M^2_\lambda}{\Lambda^2+M^2_\lambda}.
\end{equation}
The dimensionful mass eigenvalues (no tilde) 
$M^2_\lambda$, with $\lambda=\pm$, are given in 
Eq.~(\ref{defD}), and the exponents are
\begin{align}
\label{abc}
\alpha_{nm}=&\frac{\alpha_2\,\beta^2}4(n^2+m^2),\nonumber\\
\beta_{nm}=&\frac{\alpha_2\,\beta^2(M_2^2-M_1^2)(m^2-n^2)}{8D},
\nonumber\\
\gamma_{nm}=& \frac{\alpha_2\,\beta^2 n m J}{2D}\,.
\end{align}
The exponents are constant under the RG flow (they involve 
the dimensionful mass parameters which do not run). 
The quantity $D$ is defined in Eq.~(\ref{defD}),
and the flavour symmetry (which entails $M_1 = M_2$) leads to the 
corresponding symmetry $n \leftrightarrow m$ 
in Fourier space ($\beta_{nm}=0$). For flavour symmetry,
the invariance $n \leftrightarrow m$ is preserved under the 
RG flow. Note that $\alpha_{nm}$ should not be confused with 
$\alpha_d$ as defined in Eq.~(\ref{alphad}).
The solution for the original Fourier amplitudes is
\begin{eqnarray}
\label{mcsclaw}
\begin{pmatrix}
\tilde u_{nm}(k) \cr \tilde v_{nm}(k) 
\end{pmatrix}
&=&
\left(\frac{k}{\Lambda}\right)^{-2} \,
\biggl[\prod_{\lambda=\pm1}
[R_\lambda(k)]^{\alpha_{nm}+\lambda \beta_{nm}}  \biggr]
{\underline {\underline {\mathcal O}}}_{\,nm}
\begin{pmatrix}  
\tilde u_{nm}(\Lambda) \cr\tilde v_{nm}(\Lambda) \end{pmatrix}
\nn
\end{eqnarray}
with the transformation matrix 
\begin{equation}
{\underline {\underline {\mathcal O}}}_{\,nm} = 
\begin{pmatrix} 
\mbox{~cosh~}\delta_{nm}  &  \mbox{~sinh~} \delta_{nm}
\cr
\mbox{~sinh~} \delta_{nm}&  \mbox{~cosh~} \delta_{nm}
\end{pmatrix},
~~ 
\delta_{nm}=\gamma_{nm}\sum_{\lambda=\pm}\lambda
\ln R_\lambda (k).
\end{equation}
Equation (\ref{mcsclaw}) contains the general expression for 
the mass-corrected UV scaling law for a 2FMSG-type model.  

If we restrict the 2FMSG model
to only one nonvanishing Fourier mode $\tilde u_{01}$ 
of the periodic potential, as it is suggested by 
the structure of
the bare Lagrangian (\ref{s2fmsg}), then we see that no other modes are 
generated by the RG flow corresponding to the mass-corrected UV scaling laws,
\begin{eqnarray}\label{sol1} 
 \begin{pmatrix} {\tilde u}_{01}(k) \cr  {\tilde u}_{10}(k)\end{pmatrix}
&=&\begin{pmatrix} {\tilde u}_{01}(\Lambda)  \cr  {\tilde u}_{10} (\Lambda)
 \end{pmatrix}
\left({k\over\Lambda}\right)^{-2}
[R_+ (k)R_-(k)]^{{\alpha_2 \beta^2 \over 4}}
\biggl[\frac{R_+ (k)}{R_-(k)}\biggr]^{
 {\alpha_2 \beta^2 (M_1^2 - M_2^2) \over 8D}} \,.
\end{eqnarray}
For the S2FMSG model with the only nonvanishing couplings
${\tilde u}(k) = {\tilde u}_{01}(k)= {\tilde u}_{10}(k)$,
the scaling laws reduce to
\begin{eqnarray}\label{sol1sym} 
{\tilde u}(k)
&=& {\tilde u}(\Lambda) 
\left({k\over\Lambda}\right)^{-2}
[R_+ (k)R_-(k)]^{{\alpha_2 \beta^2 \over 4} } .
\end{eqnarray}

We now specialize to the LSG model,
inserting one vanishing mass eigenvalue $M_-^2=0$,
and using $M_+^2 > 0$, to obtain
\begin{eqnarray}\label{sol1lsg} 
{\tilde u}(k) 
&=& {\tilde u}(\Lambda) 
 \left({k\over\Lambda}\right)^{-2+{\hf \alpha_2 \beta^2} }
[R_+ (k)]^{{\alpha_2 \beta^2 \over 4} } .
\end{eqnarray}
Finally, for the ML2FSG model with two vanishing 
mass eigenvalues, one recovers the particular case of 
Eq.~(\ref{linsol}),
\begin{eqnarray}\label{sol1mlessg} 
\begin{pmatrix} {\tilde u}_{01}(k) \cr  {\tilde u}_{10}(k)\end{pmatrix}
&=&
\begin{pmatrix} {\tilde u}_{01}(\Lambda)  \cr  {\tilde u}_{10} (\Lambda)
\end{pmatrix}
\left({k\over\Lambda}\right)^{-2+{ \alpha_2 \beta^2} } \,,
\end{eqnarray}
without any mass corrections.

We now discuss the consequences of the mass-corrected UV scaling laws 
(\ref{mcsclaw}) for the particular cases as listed in
Eqs.~(\ref{sol1})---(\ref{sol1mlessg}).
For the general (S)2FMSG model with positive definite mass matrix,
we find that according to Eq.~(\ref{mcsclaw}), 
there is no Coleman-type fixed point irrespective 
of the value of the parameter $\beta$. 

A Coleman-type fixed point can in principle 
only be obtained for models where one or both 
of the mass eigenvalues vanish, as it is 
the case for for the LSG and the ML2FSG models.
Having transformed the mass matrix to diagonal form by an
appropriate global rotation in the internal space, these models 
exhibit explicit periodicity in one or both of the 
independent orthogonal directions in 
the internal space. According to Eq.~(\ref{sol1}),
an expression of the structure $(k/\Lambda)^{-2+\eta}$,
with $\eta$ depending on $n$, $m$, and $\beta$,
appears in the UV scaling laws if and only if at least one mass
eigenvalue vanishes. The term
$(k/\Lambda)^{-2+\eta}$ starts to dominate the flow of the 
couplings when $k$ approaches the scale $M_+$. If one extrapolates the 
UV scaling laws toward the IR region, a Coleman-type fixed
point is predicted for $\eta=2$, i.e. for some critical 
value~$\beta^2 = \beta_c^2$.
A positive definite mass matrix corresponds to breaking periodicity in both
independent orthogonal directions of the internal space and results in
the removal of the Coleman fixed point, as compared to the massless case
(unbroken periodicity). 

For the LSG model with a single nonvanishing 
mass eigenvalue $M_+^2\not=0$, periodicity is broken only in a single 
direction of the internal space, and this results in the shift of the 
Coleman fixed point lying at $\beta_c^2=8\pi$ (for the massless case) to 
$\beta_c^2=16\pi$, as shown explicitly 
below. A similar fixed point has been found for the massless
one-flavour sine-Gordon model \cite{sg2,sg3}.
For the one-flavour massive sine-Gordon model, this 
fixed point disappears, as we shall discuss below. 
In general, the increasing number of flavours 
opens various ways of breaking periodicity explicitly in 
a subspace of the internal space, and this affects the existence and the 
position of the Coleman fixed point.

\subsubsection{S2FMSG Model}
\label{flowslsg}

For symmetric initial conditions at the UV scale $\Lambda$, 
the relation $\tilde u = {\tilde u}_{01} = {\tilde u}_{10}$ holds 
throughout the evolution, and Eq.~(\ref{sol1sym}) can be recast into the form
\begin{eqnarray}\label{sol2} 
 {\tilde u}(k)
  &=& {\tilde u}(\Lambda) 
\left({k\over\Lambda}\right)^{-2}   
\left({(k^2 + M^2)^2  - J^2 \over (\Lambda^2 + M^2)^2 - J^2}
\right)^{\alpha_2 \beta^2/4}.
\end{eqnarray}
We recognize immediately that for $k \to \infty$ (i.e,
$k \sim \Lambda$), this flow is 
equivalent to the massless flow (\ref{sol1mlessg}),
and that the corrections to the massless flow are of order
$M^2/k^2$, and $J^2/k^2$, as it should be (based on dimensional 
arguments, and because the corrections have to 
vanish as $k \to \infty$). It is reassuring to observe that 
the solution~(\ref{sol2}) is also consistent
with the UV scaling law (\ref{linsol})
of the symmetric massless ML2FSG model for general $n$ and $m$.
For scales $k$ approaching the mass $M_+$, however, the Fourier amplitude 
${\tilde u}(k)$ becomes relevant, independent of the choice of $\beta^2$. 
This is a very important modification of the linearized result 
in Eqs.~(\ref{linsol}) and~(\ref{sol1mlessg}):
not only is the Coleman fixed point is gone, but 
the mass-corrected flow (\ref{sol2}) also 
suggests the existence of a cross-over region where the UV irrelevant 
coupling ${\tilde u}$ turns to a relevant one. 
One thus expects the existence of a single phase for 
the general S2FMSG model with two nonvanishing 
eigenvalues of the mass matrix.

\subsubsection{ LSG model}
\label{flowlsg}

\begin{figure}[htb]
\includegraphics[width=14cm]{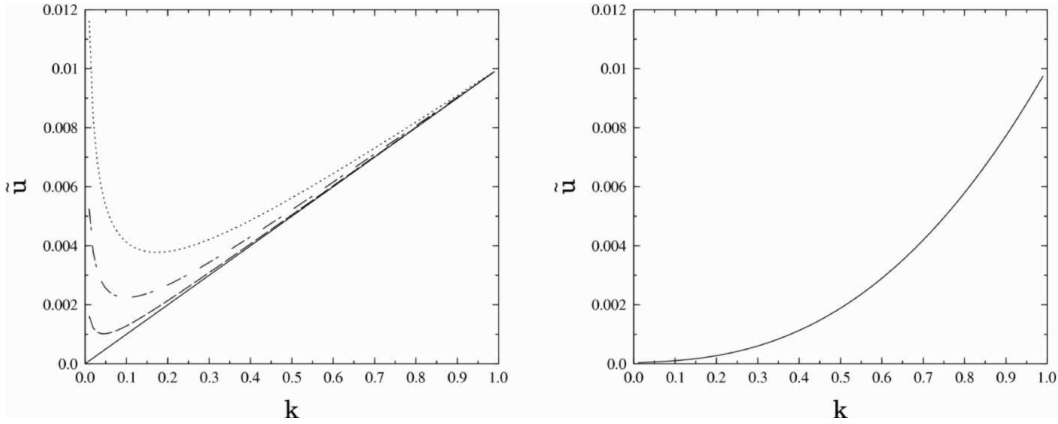}
\caption{Scaling of the dimensionless coupling constant $\tilde u $ for 
$\beta^2=12\pi$ (to the left) and for $\beta^2 = 18\pi$ (to the right), 
according to Eq.~(\ref{sol3}), for the LSG model. In the figure to the 
left, the solid line represents the UV scaling law obtained according 
to Eq.~(\ref{linsol}), and the dashed, dashed-dotted and the dotted 
lines illustrate the mass-corrected UV scaling laws  for various 
values of $J=0.002, \, 0.01, \, 0.03$, respectively. For the 
computations, the UV scale has been chosen as $\Lambda=1$.
\label{lsgfig}}
\end{figure}

We recall the mass-corrected solution (\ref{sol1lsg}), 
which is equivalent to Eqs.~(\ref{sol1sym}) and~(\ref{sol2}) for the case
$J = M$,
\begin{eqnarray}\label{sol3} 
{\tilde u}(k)
&=&
{\tilde u}(\Lambda)
\left({k\over\Lambda}\right)^{-2 + \alpha_2 \beta^2 /2}   
\left({k^2 + 2J\over \Lambda^2 + 2J}\right)^{\alpha_2 \beta^2 /4}.
\end{eqnarray}
A graphical representation can be found 
in Fig.~\ref{lsgfig}. For $8\pi < \beta^2 < 16\pi$,
the solution for $ {\tilde u}$ has a minimum at 
$k_{min} = [J (4-\alpha_2\beta^2 )/( \alpha_2\beta^2 -2)]^{1/2}$.  

If $\beta^2>\beta^2_{c} =16\pi$, the Fourier amplitude $\tilde u$ remains 
an irrelevant coupling constant even in the IR region. This suggests
that the LSG model may exhibit two phases, separated by the Coleman 
fixed point. The coupling $u$, which 
plays the role of the fugacity of 
the layered vortex gas has a completely different
behaviour in these two phases.
The critical value (critical temperature) for the 
layered system  $\beta_c^2 =16\pi$ persists;
this critical value holds irrespective of the 
mass eigenvalue $M_+^2=2J$, the only criterium 
being that $M_+^2$ should be nonvanishing.

By contrast, if we set $J=0$ explicitly,
we arrive at the symmetric massless
ML2FSG model with the critical value  $\beta_c^2 =8\pi$ 
[see Eq.~(\ref{sol1mlessg})]. The limit $J\to 0$ is in 
that sense nonuniform, and the phase structure is 
also nonuniform, because 
an entire symmetry gets restored for $J=0$ (periodicity 
in both directions of the internal space).

\begin{figure}[htb]
\includegraphics[width=8cm]{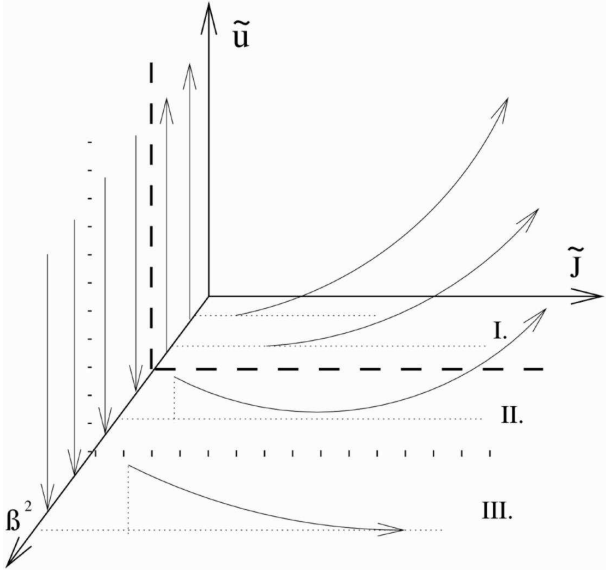}
\caption{Phase diagram of the LSG model based on 
the mass-corrected UV scaling law (\ref{sol3}). As there is no 
evolution for $\beta^2$ in $d=2$ in the LPA, the RG trajectories lie 
in planes of constant~$\beta^2$. The arrows indicate the direction of 
the flow ($k\to 0$) in which the dimensionless mass eigenvalue 
$2{\tilde J}_k = k^{-2} \, 2J$ increases. 
In the $(\tilde u,~\beta^2)$ plane, the phase diagram of 
the ML2FSG model ($\tilde J=0$) is depicted where 
the dashed line at $\beta_c^2 = 8\pi$ separates the two phases.
For the LSG, one finds two phases separated 
by the plane at  $\beta_c^2 = 16 \pi$  (indicated by the dotted lines).
In the phase with $\beta^2<  16 \pi$, two (sub-)regions can be recognized.
In region I, the trajectories have the same tendency as for $J = 0$: 
in particular, ${\tilde u} $ remains a relevant (increasing) 
parameter for $k\to 0$. In region II, the UV irrelevant (decreasing) 
$\tilde u$ becomes a relevant (increasing) parameter after a cross-over 
region. In the phase with  $\beta^2>  16 \pi$ (region III), the 
Fourier amplitude ${\tilde u}$ remains irrelevant 
during the RG flow.
\label{lsgphase}}
\end{figure}

For the LSG model, a preliminary phase diagram,
as suggested by the mass-corrected flow, is plotted in 
Fig.~\ref{lsgphase}. To this end, we have to assume 
that the mass-corrected UV scaling law (\ref{sol3}) holds 
at least qualitatively in the IR region. 
This conjecture is supported by numerical calculations,
based on the nonlinear terms $\cF_2({\tilde U}_k)$
in Eq.~(\ref{uflex}), as described 
below in Sec.~\ref{nonlinflowlsg}. 
Preliminary numerical results, based on the 
full WH RG equation (\ref{Uflow}) which goes beyond 
the subleading nonlinear term analyzed in Sec.~\ref{nonlinflowlsg}, 
also support this conjecture 
(the latter calculations will be presented in detail elsewhere).

For the LSG, the broken periodicity in one 
direction of the internal space leads to 
\begin{itemize}
\item[-] the existence of two phases with different IR fixed points, 
${\tilde u} \to \infty$ for $\beta^2<\beta_c^2$ and
${\tilde u} \to 0$ for  $\beta^2>\beta_c^2$, respectively,  and
\item[-] an intermediate
region in the phase diagram where the UV 
irrelevant vortex fugacity ${\tilde u}$ 
becomes relevant in the IR scaling regime, after passing a cross-over
regime.
\end{itemize} 
In Fig.~\ref{lsgfig} (regions I and III), 
the overall scaling behaviour of the vortex 
fugacity is the same as that for the symmetric ML2FSG model, 
and in particular, no cross-over 
regime appears in the flow of $\tilde u$. 
The cross-over regime will be of particular interest for 
further numerical calculations, based on the full WH RG equation 
(\ref{Uflow}).
 
\subsubsection{MSG model}
\label{flowmsg}

\begin{figure}
\includegraphics[width=8cm]{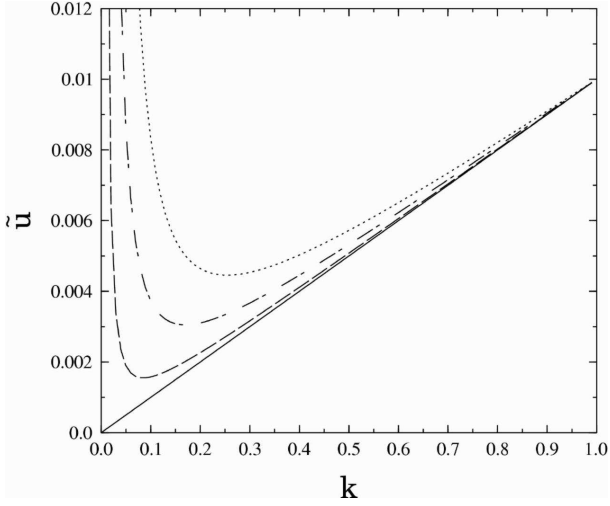}
\caption{Scaling of the dimensionless coupling constant $\tilde u $
of the MSG model for $\beta^2 = 12 \pi$. The solid line represents 
the UV scaling law (\ref{linsol}) for the massless SG model. 
The dashed, dashed-dotted and the dotted lines depict the 
mass-corrected UV scaling laws (\ref{sol4}) for the MSG model,
for various values of 
$M^2=0.0036, \, 0.0144, \, 0.0324$, respectively.  
In the IR, the mass-corrected RG flow is drastically and qualitatively 
different from the massless flow, even for small mass parameters,
due to the broken internal symmetry.
\label{msgfig}}
\end{figure}

It is enlightening to discuss the mass-corrected UV scaling laws for 
the (one-flavour) MSG model, another particular case  with entire breaking 
of periodicity in the internal space. Formally, the UV scaling laws for 
the MSG model can be obtained  from Eq. (\ref{mcsclaw}) by setting 
$M_1^2=M^2$, $M_2^2=J=0$, which implies that $D=M^2/2$ in 
Eq.~(\ref{abc}). In this case, flavour symmetry would be broken,
but the two flavours actually decouple, and thus we restrict
the discussion to a single flavour. We also restrict ourselves to a
single Fourier mode in the blocked potential with $(n=1,m=0)$ and the 
amplitude $\tilde u = {\tilde u}_{10}$. The 
UV mass-corrected RG evolution reads
\begin{eqnarray}\label{sol4} 
{\tilde u}(k) &=& {\tilde u}(\Lambda)   
\left({k\over\Lambda}\right)^{-2}   
\left({k^2 + M^2 \over \Lambda^2 + M^2}\right)^{\alpha_2 \beta^2 /2}. 
\end{eqnarray}
This reproduces  the  UV behaviour (\ref{linsol}) of the corresponding 
massless model for scales $M \ll k \sim \Lambda $, where ${\tilde u}(k)$ is  
irrelevant (relevant) for $\beta^2 > 8\pi$ $~(<8\pi)$. However, the 
mass-corrected UV scaling law (\ref{sol4}) of 
the MSG model to the IR limit predicts a cross-over at scales 
$k^2 \sim \ord{M^2}$ (even) for $\beta^2 > 8\pi$ below which the coupling 
${\tilde u}(k)$ becomes relevant (see Fig. \ref{msgfig}). Thus, 
irrespective of the choice of $\beta^2$, the coupling ${\tilde u}(k)$ 
is suggested to be IR relevant according to the (extrapolation of) the 
mass-corrected UV scaling law (\ref{sol4}) into the IR region.
 
The mass-corrected UV scaling law in Eq.~(\ref{sol4}) accounts for the 
explicit breaking of periodicity in the (one-dimensional) internal 
space via the nonvanishing mass term and results in the removal of 
the Coleman fixed point, as compared to the massless case. 

\begin{figure}
\includegraphics[width=8cm]{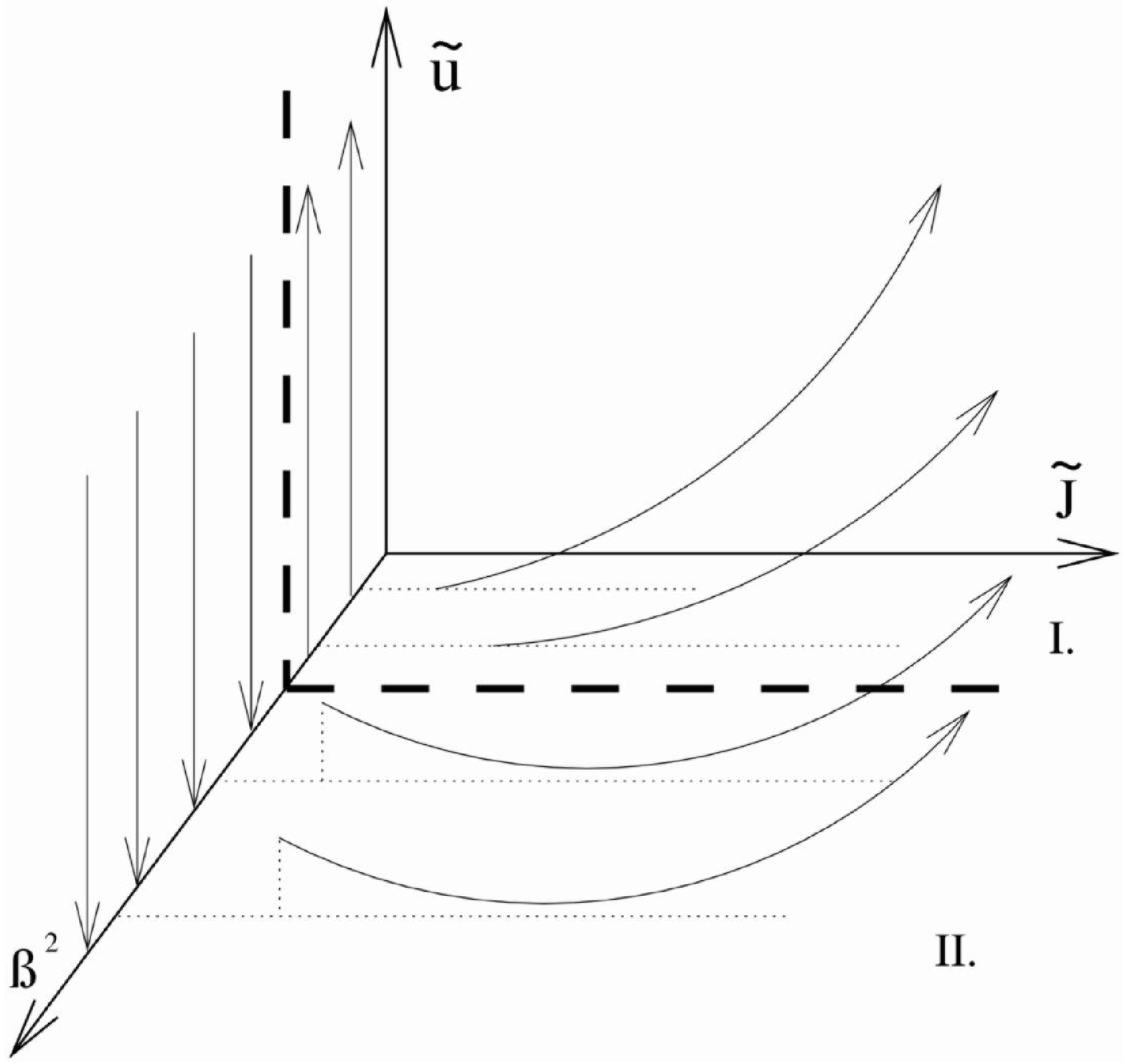}
\caption{Schematic phase structure of the MSG model based on the analytic
solution (\ref{sol4}).
As in Fig.~\ref{lsgphase}, the results are obtained in the local-potential
approximation, where there is no evolution for $\beta^2$ and the
RG trajectories are always parallel to the $\tilde M^2 = \tilde J$ axis.
The arrows indicate the direction of the RG flow ($k\to 0$). 
The WH-RG equation
(\ref{WHdim}) gives a trivial scaling for the coupling
${\tilde M^2}(k)= {\tilde J}(k) \propto k^{-2}$ 
[see Eq.~(\ref{treelevel})],
so that the mass parameters remain relevant couplings
during the whole RG flow. The $\tilde u$-$\beta^2$
plane corresponds to the phase diagram of the
massless SG model ($\tilde M^2 = \tilde
J=0$). The dashed line separates the two phases of the SG
(but not the MSG) model. The
linearization of 
the WH equation (\ref{WHlin1}) 
would predict the same two phases for the MSG model with
the same critical value $\beta^2 = 8\pi$. 
However, the mass-corrected RG treatment modifies
this picture and shows only one phase for the MSG model.
In region I, the trajectories have the same tendency as in
the massless theory; $\tilde u \equiv {\tilde u}_{01}$ is a relevant
(increasing) parameter in the UV and in the IR domain as well. In region
II, the UV irrelevant (decreasing) $\tilde u$
becomes a relevant (increasing) parameter in the IR limit,
after a crossover region, according to Eq.~(\ref{sol4}).
\label{msgphase}}
\end{figure}

\begin{figure}
\includegraphics[width=14cm]{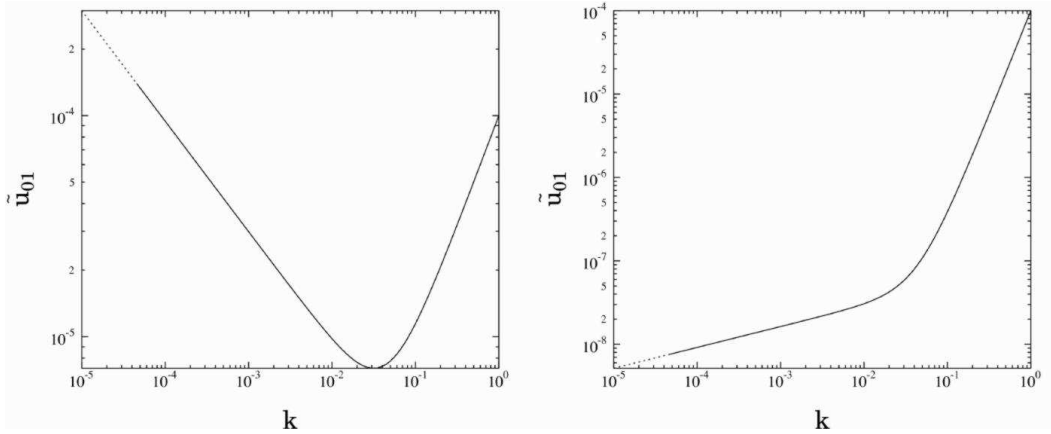}
\caption{The scaling of the dimensionless coupling constant $\tilde u_{01}$
of the LSG model is represented graphically for two different 
temperature parameters
$\beta^2 = 12 \pi$ (left) and $\beta^2 = 18 \pi$ (right).
The interlayer coupling is $J=0.001$ in both cases.
The dotted line represents the solution according to 
Eqs~(\ref{sol1lsg}) and~(\ref{sol3}), which is obtained by considering the 
linear term $\cF_1(\tilde U_k)$ in Eq.~(\ref{uflex}).
The solid line shows the solution of the RG flow 
including [in addition to 
$\cF_1(\tilde U_k)$] also 
the nonlinear term $\cF_2(\tilde U_k)$ in Eq.~(\ref{uflex}),
which leads to the system of equations (\ref{nonlinlsg}).
Both curves almost overlap, which demonstrates that 
the flow of the fundamental coupling $\tilde u_{01}$
is almost independent of the nonlinear corrections
mediated by the $\cF_2$ term. 
\label{nonlinfig1}}
\end{figure}

\begin{figure}
\includegraphics[width=14cm]{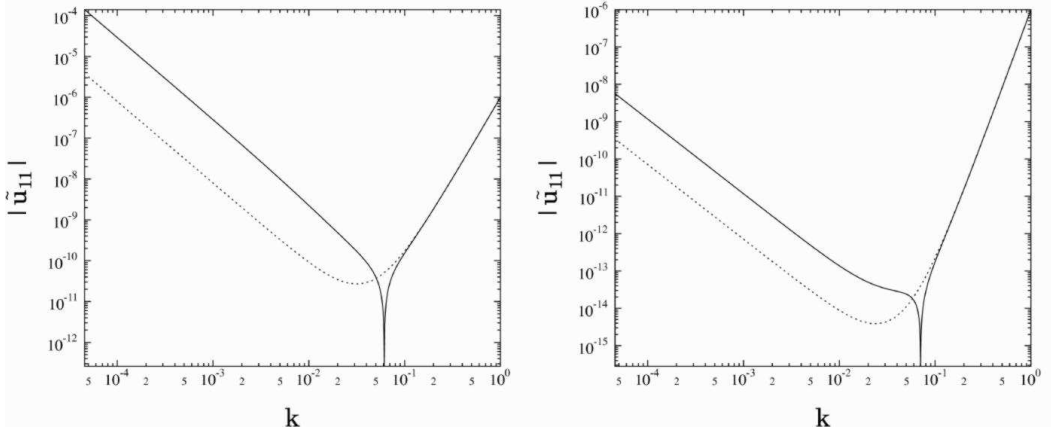}
\caption{The scaling of the dimensionless coupling constant 
$\vert\tilde u_{11}\vert$ (``higher harmonic'') 
of the LSG model is shown for $\beta^2 = 12 \pi$ 
(left) and $\beta^2 = 18 \pi$ (right) and $J=0.001$. The 
solid and dotted curves are obtained with and 
without the nonlinear terms, as in Fig.~(\ref{nonlinfig1}),
but for a different coupling parameter ($\tilde u_{11}$
instead of $\tilde u_{01}$), and with an initial
condition $\tilde u_{11} (\Lambda) = 10^{-4}$ at the 
UV scale $\Lambda = 1$.  The solution for $\tilde u_{11}$,
including the nonlinear terms [see Eq.~(\ref{nonlinlsg})],
changes sign near $k \approx 7 \times 10^{-2}$
(so that $\ln |\tilde u_{11}| \to -\infty$), whereas the 
flow with linear mass corrections
predicts no change of sign (dotted line).
\label{nonlinfig2}}
\end{figure}

\begin{figure}
\includegraphics[width=14cm]{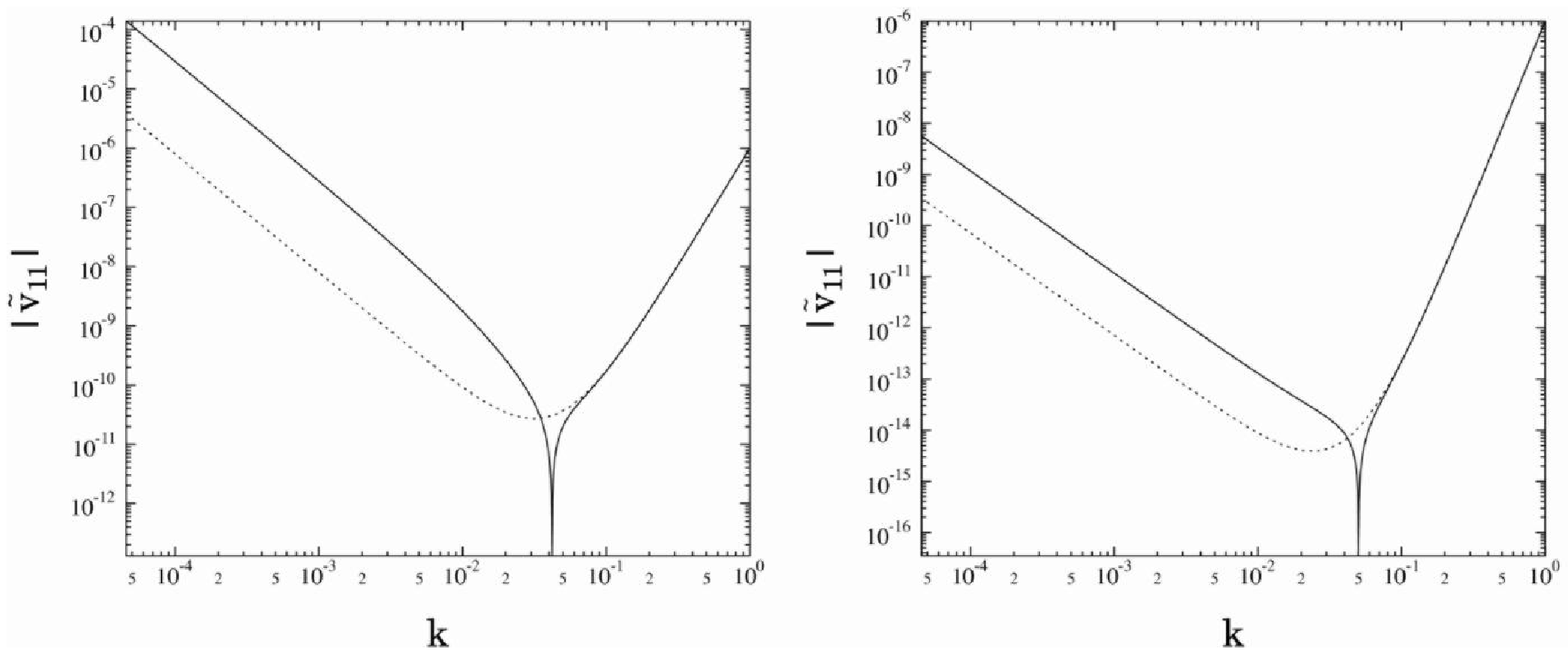}
\caption{\label{nonlinfig3} 
The same as Fig. \ref{nonlinfig2} for the dimensionless coupling
constant $\vert\tilde v_{11}\vert$ (LSG model).
In the UV, the two solutions with and without nonlinear terms
overlap. In the IR, the two solutions appear to follow
similar scaling laws, with approximately equal double-logarithmic
derivatives $\partial\ln \vert {\tilde v}_{11}(k) \vert/\partial\ln k$.}
\end{figure}

\subsection{Extended UV scaling laws for the LSG model}
\label{nonlinflowlsg}

In Secs.~\ref{flowslsg},~\ref{flowlsg}, and~\ref{flowmsg},
we restricted the discussion to the linear corrections 
${\cF}_1(\tilde U_k)$ as listed in Eq.~(\ref{uflex}).
Here we investigate a further modification of the UV scaling laws
toward the lower scales, by taking into account the nonlinear term 
${\cF}_2(\tilde U_k)$ quadratic in the potential on the right hand side of 
Eq. (\ref{uflex}). For the sake of simplicity, we restrict ourselves 
to the LSG model. We would like to demonstrate that the
nonlinear term ${\cF}_2(\tilde U_k)$
(i) does not change the phase structure obtained
on the basis of the mass-corrected UV scaling law (\ref{mcsclaw}), 
but (ii) may have a significant effect on the effective potential
obtained for $k \to 0$. 
Thus, one is inclined to suggest that the mass-corrected UV scaling 
laws enable one to obtain the correct phase structure, although the 
nonlinearities as implied by the 
full WH equation (\ref{Uflow}) play a decisive role in the cross-over region,
and for a detailed quantitative analysis of the IR region
and the effective potential. 

Equating the coefficients of the corresponding Fourier modes on the both 
sides of Eq. (\ref{uflex}), one arrives at the set of equations for the 
scale-dependent Fourier amplitudes. For the first few Fourier amplitudes
$\tilde u_{01}=\tilde u_{10}$, $\tilde u_{11}$ and 
$\tilde v_{11}$, the nonlinear RG equations read
\begin{subequations} 
\label{nonlinlsg}
\begin{eqnarray} 
\label{nonlinlsga}
(2+k \, \partial_k) \, \tilde u_{01} &=& 
\alpha_2 \, \beta^2 {\bf F} \, \tilde u_{01} 
\nonumber\\
& & + \alpha_2 \, \beta^4 
\left[\left({{\bf F}^2 \over 2} + {\bf G}^2 \right) 
\tilde u_{01} \, \tilde u_{11} - 
2 {\bf F} \, {\bf G} \, \tilde u_{01} \, \tilde v_{11}   
\right], \\
\label{nonlinlsgb}
(2+k \, \partial_k) \, \tilde u_{11} &=& 
\alpha_2 \, \beta^2 
\left[2 {\bf F} \, \tilde u_{11} - 2{\bf G} \, \tilde v_{11} \right]
+ \alpha_2 \, \beta^4 
\left[{\bf G}^2 \, \tilde u_{01}^2 \right], \\
\label{nonlinlsgc}
(2+k \, \partial_k) \, \tilde v_{11} &=& 
\alpha_2 \, \beta^2 
\left[2 {\bf F} \, \tilde v_{11} - 2{\bf G} \, \tilde u_{11} \right],
\end{eqnarray}
\end{subequations} 
using the notations
\begin{eqnarray}
{\bf F} = {k^2+  J \over k^2+2 J}, \hspace{1cm}
{\bf G} = { J \over k^2+2 J}.
\end{eqnarray}
The nonlinear terms generate ``higher harmonics.''
Specifically, we have the situation that 
even for vanishing initial values of the 
couplings of the higher-order Fourier modes at the UV scale $\Lambda$, 
their nonvanishing values are
generated by the fundamental modes $(1,0)$ and $(0,1)$ due to the nonlinear 
term proportional 
$\tilde u_{01}^2$, which can be found on the right hand side of 
Eqs. (\ref{nonlinlsgb}). Higher-order Fourier modes with nonvanishing
couplings appear in general during the blocking of the LSG model due to the 
nonlinearities incorporated in the logarithm on the right hand side of
Eq. (\ref{Uflow}). The general ansatz (\ref{cl}) for the blocked potential
was motivated by this mixing of the modes and by symmetry 
considerations.

According to Eq. (\ref{sol3}), the coupling
$\tilde u_{01}(k)$ decreases monotonically with decreasing scale $k$, but
its logarithmic slope $\partial\ln\tilde u_{01}(k)/\partial\ln k$
is predicted to change from
$-2+\alpha_2\beta^2$ for $J\ll k^2 <\Lambda^2$ to  $-2+\alpha_2\beta^2/2$
for $ k^2 \ll J$.
The couplings of the higher harmonics should be
irrelevant in the UV: both
$|\tilde u_{11}(k)|$, and
$|\tilde v_{11}(k)|$ should be proportional to
$k^{-2+2\alpha_2\beta^2}$.
Equation (\ref{sol3}) also predicts
that $|\tilde u_{11}(k)|$, and
$|\tilde v_{11}(k)|$ should
become relevant in the IR region, following essentially the tree-level
scaling $\sim k^{-2}$.

As shown in Figs.~\ref{nonlinfig1}---\ref{nonlinfig3},
these basic features are not modified by the
nonlinear terms. Numerical solutions of Eq. (\ref{nonlinlsg}) are found
for initial conditions which are chosen so that
$|\tilde u_{01}(\Lambda)|\gg |\tilde u_{11}(\Lambda)|$
and $|\tilde u_{01}(\Lambda)|\gg |\tilde v_{11}(\Lambda)|$
at the UV scale, and $\beta^2$ assumes the values of
$12\pi$ and $18\pi$ 
(see Figs.~\ref{nonlinfig1}---\ref{nonlinfig3}). 
The scaling of the fundamental modes $\tilde u_{01}(k)$ 
is only marginally influenced by the nonlinear terms 
(Fig. \ref{nonlinfig1}). 
The situation is somewhat different for 
$\tilde u_{11}(k)$ and
$\tilde v_{11}(k)$. If the nonlinear terms
are added, then the couplings $\tilde u_{11}(k)$ and 
$\tilde v_{11}(k)$ change sign in the cross-over region. 
The flow diagrams reflect the same phase structure as 
obtained on the basis of the mass-corrected UV scaling laws. 
In particular, 
the fact that the couplings $\tilde u_{11}(k)$ and $\tilde v_{11}(k)$ follow 
the tree-level scaling in the IR region ($\propto k^{-2}$) 
means that the dimensionful 
couplings (obtained via multiplication by $k^2$) 
tend to nonvanishing finite constants in the limit $k\to 0$. 
For $\beta^2< \beta_c^2$, the fundamental 
dimensionful coupling $ u_{01}$ behaves 
similarly, whereas for $\beta^2> \beta_c^2$ it tends to zero. Thus, one 
expects---in both phases---a nonvanishing periodic piece of the 
effective potential, as opposed to the massless SG model when the 
periodic effective potential should be a trivial constant due to the
 requirement of convexity \cite{sg2,sg3}.

\section{Summary}
\label{conclu}

The differential renormalization group (RG) in momentum space with a
sharp cut-off (Wegner's and Houghton's method) has been applied in the local 
potential approximation (LPA) to a general two-flavour massive sine-Gordon 
(2FMSG) model, as defined in Sec.~\ref{def}. 
The ansatz used for the blocked potential contains a 
mass term and a contribution 
which is periodic in the different directions of the 
internal space [see Eq.~(\ref{def1})]. 
The bare Lagrangians under study have only one nonvanishing 
Fourier mode [see Eq.~(\ref{genstruc})].
Particular attention has been paid to the layered 
sine-Gordon (LSG) model, as defined in 
Eq.~(\ref{lsg}), which is the bosonized version of the 
multi-flavour Schwinger model.
In general, we consider models with two flavours
(two interacting scalar quantum fields) with an
interaction periodic in the internal space spanned 
by the field variables.

For the massive SG-type models, the usual perturbative approach to 
renormalization is not applicable. One should preserve the symmetry of 
the periodic part keeping the Taylor expansion of the potential intact.
``Polynomial'' self-interactions proportional 
to $\phi^{n}$, obtained by the Taylor 
expansion of the periodic potential,
should be summed up and considered as one 
composite operator [which might be of the form $\cos(\beta \phi)$]. 
This can only be achieved in the framework of non-perturbative 
renormalization group methods.

It has been shown that the dimensionful mass matrix remains constant in 
the LPA, under the RG flow. 
The explicit breaking of the periodicity by
mass terms modifies the properties of the 
scaling laws and the periodic blocked potential significantly.
UV scaling laws for the massless SG models exhibit a Coleman fixed point. 
For massive models, the determination of the 
UV scaling laws has to include mass corrections (see Sec.~\ref{rgflow}).
When periodicity is partially broken, with one 
nonvanishing mass eigenvalue, the Coleman fixed point
is found to be shifted.
With an entirely broken periodicity,
we find a complete disappearance of the Coleman fixed 
point. 

For the particular case of the 
LSG model, periodicity is only partially 
broken, and the existence of two phases is suggested by the 
RG flow. The fundamental mode ${\tilde u}_{01}$ 
of the periodic potential is irrelevant 
and relevant in the IR scaling region,
depending on whether $\beta^2 > 16\,\pi$ or 
$\beta^2 < 16\,\pi$, respectively.
The RG flow of the UV irrelevant amplitude of the fundamental mode 
may pass a cross-over region 
($8\,\pi < \beta^2 < 16\,\pi$),
before becoming relevant in the IR regime.
The mass-corrected RG flow 
is beyond the ``dilute gas approximation''
which would correspond to the flow given by
Eq.~(\ref{WHlin1}).

In view of our analysis of the S2FMSG (Sec.~\ref{flowslsg}),
of the LSG (Secs.~\ref{flowlsg} and~\ref{nonlinflowlsg}) 
and the MSG model (Sec.~\ref{flowmsg}), we may 
suggest that the Coleman fixed point disappears, when periodicity is 
explicitly broken by mass terms in both independent directions of 
the internal space. Thus, one expects the existence of a single phase for 
the MSG model (see Fig. \ref{msgphase}). Of course,
a final and definite conclusion would require
a full numerical solution of the flow  
equation (\ref{Uflow}) for these models. However, 
we are in the position to remark that preliminary 
numerical results appear
to support the results based on the mass-corrected UV RG flow,
as reported in the current article.
The interesting cross-over region, as shown in
Figs.~\ref{lsgphase} and~\ref{msgphase}, suggests that the numerical 
determination of the effective potential can provide operators, which 
are relevant for IR physics although they are irrelevant at the
UV scale.

The subleading nonlinear terms in RG flow have been analyzed 
in Sec.~\ref{nonlinflowlsg}, which is a step toward the 
full solution of the WH equation (\ref{Uflow}).
The nonlinear terms are quadratic in the periodic blocked potential.
Due to the nonlinearity of the flow, higher order 
Fourier modes, normally suppressed at the UV cut-off, appear in the periodic 
blocked potential. For the LSG model,
it has been demonstrated that the 
quadratic nonlinear terms play a negligible role for the 
RG evolution of the fundamental coupling ${\tilde u}_{01}$,
provided the higher harmonics are suppressed at the UV scale
(as it should be in view of the given structure of the 
bare Lagrangians). However, the 
nonlinear terms play an important role in the behaviour of the 
UV irrelevant couplings of the higher harmonics 
in the cross-over region.
 
Another rather surprising aspect concerns the 
structure of the effective potential for theories 
with a nonvanishing mass matrix as opposed to their 
massless counterparts: namely, for the 
``massive'' case, 
one expects a nonvanishing periodic of the effective potential,
as opposed to the massless SG model, where the 
simultaneous requirements of periodicity 
and convexity result in a field-independent effective potential.

%
%
\section*{Acknowledgements}

I. N\'{a}ndori thanks the Max--Planck--Institute for Nuclear 
Physics, Heidelberg, for the kind hospitality extended on the 
occasion of a guest researcher appointment in 2004 during which 
part of this work was completed. Numerical calculations were
performed on the high-performance computing facilities
of the Max--Planck--Institute, Heidelberg. I. N\'{a}ndori takes 
a great pleasure in acknowledging discussion with K. Vad, 
S. M\'esz\'aros and J. Hakl. U. D. Jentschura acknowledges 
support by the Deutsche Forschungsgemeinschaft (Heisenberg program).
S. Jentschura is acknowledged for 
carefully reading the manuscript.

\appendix

\section{Bosonization of the Multi-Flavour Schwinger Model}
 \label{2fmsm}

In this section, we dwell on the fact that the MSG model (\ref{msg}) 
and the  LSG model (\ref{lsg}) are the theories obtained by
bosonization from the massive Schwinger model (1+1 dimensional QED)
obeying $U(1)$ and $SU(2)$ global flavour symmetries, respectively.
The multi-flavour Schwinger model has not been studied as extensively 
as the massive Schwinger model, the case with $U(1)$ flavour symmetry. 
The latter proved to be interesting since it shows confinement 
properties. However, the relative ignorance toward the multi-flavour 
Schwinger model is perhaps not fully justified as it shows more 
resemblance to the 4-dimensional QCD, because the model features a 
chiral symmetry breakdown~\cite{HeHoIs1995}.

Two--dimensional QED with an $SU(2)$ internal symmetry can be 
characterized by the Lagrangian
\begin{equation}
\label{2dqed}
{\mathcal L}=\sum_{i=1,2}\bar\psi_i(\sla\partial-m-e\sla A)\psi_i
-\frac14 F_{\mu\nu}F^{\mu\nu}\,.
\end{equation}
Here $A_\mu$ is the vector potential of the photon field.
The $\psi_i$ ($i=1,2$) denote an $SU(2)$ flavour-doublet of 
fermions. Furthermore, the field-strength tensor is given by
$F_{\mu\nu}=\partial_\mu A_\nu-\partial_\nu A_\mu$, and $m$ and 
$e$ are the bare rest mass of the electron and  the bare coupling 
constant, respectively. The model (\ref{2dqed}) was shown to be 
capable~\cite{Fi1979} of describing materials with a zero net 
charge, but with a non-zero flavour charge, interpreted as 
`baryon number' density, a kind of matter in neutron stars.
Bosonization of the model (\ref{2dqed}) proceeds according to 
the following rules~\cite{Co1973,Co1975,Co1976},
\begin{subequations}
\label{bosonize}
\begin{eqnarray}
:\bar\psi_i\psi_i:&\to& -cmM\cos(2\sqrt{\pi}\phi_i),\\
:\bar\psi_i\gamma_5\psi_i:&\to& -cmM\sin(2\sqrt{\pi}\phi_i), \\
:\bar\psi_i\gamma_\mu\psi_i: &\to&
 \frac1{\sqrt{\pi}}\varepsilon_{\mu\nu}\partial^\nu \phi_i,\\
:\bar\psi_i i \sla\partial\psi_i: &\to& \frac12 N_m (\partial\phi_i)^2,
\end{eqnarray}
\end{subequations}
where $i=1,2$, and there is no sum on $i$. Here, $N_m$ denotes 
normal ordering with respect to the fermion mass $m$,
and $c=\exp{(\gamma)}/2\pi$ with the Euler constant $\gamma$. 
In the case of an equal mass and opposite charges of the
two fermions, the bosonized form of the theory becomes
\begin{eqnarray}
\label{msmb}
{\mathcal H}&=&N_m\biggl[\frac12\Pi_1^2+\frac12\Pi_2^2
+\frac12(\partial_1\phi_1)^2+\frac12(\partial_1\phi_2)^2\nonu
&&-cm^2\cos(2\sqrt\pi\phi_1)-cm^2\cos(2\sqrt\pi\phi_2)
-\frac{e^2}{2\pi}(\phi_1-\phi_2)^2 \biggr]\,.
\end{eqnarray}
The theory defined by the Hamiltonian (\ref{msmb}) is identical 
to the LSG model (\ref{lsg}) under an appropriate identification 
of the coupling constants of the two models ($\beta^2 = 4\pi$).

%
%
\section{Some notes on the Wegner-Houghton equation}
\label{WHapp}

As has already been mentioned in Sec.~\ref{nonpert},
the WH-RG equation has to be projected into a particular 
functional subspace, in order to reduce the search for a functional
(the blocked action) to the calculation of
an appropriate function. Here, we assume that the
blocked action contains only local interactions. We 
use the approach outlined in~\cite{janosRG,ZJRG}, expand it in 
powers of the gradients of the fields $\phi_1$ and $ \phi_2$,
and keep only the leading-order terms; thus we arrive at an
ansatz for the blocked action. Indeed, for the $d=2$ LSG-type models
with two scalar fields $\phi_1$ and $\phi_2$,
the blocked action reads
\begin{equation}
\label{eucac}
S_k = \, \int {\mathrm d}^2 x \left[
 {1\over2} \, (\partial\phi_1)^2 + {1\over2} 
(\partial\phi_2)^2 \, + \, V_k (\phi_1,\phi_2) \right]\,.
\end{equation}
The evolution of the blocked potential $V_k$ in the direction of
decreasing $k$ is supposed to be satisfying the
following generalized WH-RG equation for two interacting 
fields in $d=2$,
\begin{equation}
\label{WH}
k \, \partial_k V_k = -{k^2\over 4\pi}
\ln \left({[k^2 + V^{11}_k] [k^2 + V^{22}_k] -
[V^{12}_k]^2  \over k^4} \right)\,,
\end{equation}
where 
\begin{equation}
V^{ij}_k \equiv \partial_{\phi_i}\partial_{\phi_j}V_k\,.
\end{equation}
We recall that $V_k$ is a function of functions $\phi_i$,
so that the differentiations with respect to the
$\phi_i$ and to the $k$ need to be carefully distinguished.
The equation (\ref{WH}) is nonperturbative as it does not imply
an expansion of $V_k$ in powers of its arguments $\phi_1$
and $\phi_2$.  The derivation of the (generalized) WH equation (\ref{WH})
for two-component models
has been inspired by techniques outlined for ${\mathcal O}(N)$-symmetric
models~\cite{EyMoNiZJ1996}.

One actually has a certain freedom in constructing the WH
equation, which becomes apparent when adding to the
Euclidean action in (\ref{eucac}) a field-independent
term. This freedom generates a class of WH equations characterized
by the structure
\begin{equation}
\label{WHgen} 
k \, \partial_k V_k = -{k^2\over 4\pi}
\ln \left({[k^2 + V^{11}_k] [k^2 + V^{22}_k] -
[V^{12}_k]^2  \over f(k)} \right)\,,
\end{equation}
with the requirement that ${\rm dim} f(k) = {\rm dim}\,k^4$,
and this freedom gives us the possibility to 
discard the term $\ln(1+ {\tilde \mu}^2)$ on the right hand
side of (\ref{logexpand}).
The WH-RG equation (\ref{WH}), rewritten in terms of dimensionless
quantities, yields Eq.~(\ref{WHdim}).

The dimensionless
WH-RG equation (\ref{WHdim}) is applicable for the LSG type
models defined in Sec.~\ref{def}, and one can solve it for a particular
field-theoretical model by projecting $\tilde V_k$
onto a particular space of functions, with appropriate
UV boundary conditions for the RG evolutions. Of course,
the functional ansatz for the blocked
potential should be rich enough in order to ensure that the RG
flow does not leave the chosen subspace of blocked potentials, and
it should preserve all symmetries of the original model at the
UV cutoff scale $k=\Lambda$. For example, the blocked potential
for the LSG model should be invariant under the exchange of the field
variables, $\phi_1 \leftrightarrow \phi_2$ because the layers
are physically equivalent, and it should also preserve
the symmetries
$\phi_i \to -\phi_i$ and $\phi_i \to \phi_i + 2\pi/\beta$
which are present in the bare Lagrangian.
In the cases of interest for the current study, 
all these requirements
are fulfilled by the ansatz (\ref{cl}) 
for the dimensionless blocked potential.


\begin{thebibliography}{99}
\bibitem{LGZJ1980} J. C. Le Guillou, J. Zinn-Justin,
Phys. Rev. B{\bf 21} (1980) 3976.
\bibitem{GuZJ1996} R. Guida, J. Zinn-Justin,
Nucl. Phys. B{\bf 489} (1996) 626.
\bibitem{HeHoIs1995} J. E. Hetrick, Y. Hosotani, S. Iso,
Phys. Lett. B{\bf 350} (1995) 92.
\bibitem{Fi1979} W. Fischler, J. Kogut, L. Susskind,
Phys. Rev. D{\bf 19} (1979) 1188.
\bibitem{MeWa1966} N. D. Mermin, H. Wagner, 
Phys. Rev. Lett. {\bf 17} (1966) 1133.
\bibitem{KoTh1973} J. M. Kosterlitz, D. J. Thouless, 
J. Phys. C{\bf 6} (1973) 118.
\bibitem{Ko1974} J. M. Kosterlitz, J. Phys. C{\bf 7} (1974) 1046.
\bibitem{JoKaKiNe1977} J. V. Jose, L. P. Kadanoff, 
S. Kirkpatrick, D. R. Nelson, Phys. Rev. B{\bf 16} (1977) 1217.
\bibitem{GeWe2000} G. von Gersdorff, C. Wetterich, {\em Nonperturbative
renormalization flow and essential scaling for the Kosterlitz-Thouless
transitions}, e-print hep-th/0008114.
\bibitem{sg2}I. N\'andori, J. Polonyi, K. Sailer, Phys. Rev. D{\bf 63} 
(2001) 045022.; Phil. Mag. B{\bf 81} (2001) 1615.
\bibitem{ZJRG} J. Zinn-Justin, {\em Groupe de renormalisation 
fonctionnel}, 2004 (unpublished);
{\em Groupe de renormalisation fonctionnel et \'{e}quations de champs},
2004 (unpublished).
\bibitem{EyMoNiZJ1996} G. Eyal, M. Moshe, S. Nishigaki, J. Zinn-Justin, 
Nucl. Phys. B{\bf 470} (1996) 369.
\bibitem{janosRG}J. Polonyi, Central Eur. J. Phys. {\bf 1} (2004) 1;
{\em Lectures on the functional renormalization group method},
e-print hep-th/0110026.
\bibitem{Wi1971} K. G. Wilson, Phys. Rev. D{\bf 3} (1971) 1818.
\bibitem{sanyi} S. Nagy, J. Polonyi, K. Sailer, Phys. Rev. D{\bf 70} 
(2004) 105023.
\bibitem{pierson}S. W. Pierson, Phys. Rev. Lett. {\bf 74} (1995) 2359;
Phys. Rev. B{\bf 55} (1997) 14536.
\bibitem{pierson_map}S. W. Pierson, O. T. Valls,
Phys. Rev. B{\bf 49} (1994) 662.
\bibitem{pierson_lsg}S. W. Pierson, O. T. Valls,
Phys. Rev. B{\bf 45} (1992) 13076.
\bibitem{pierson_rg}S. W. Pierson, O. T. Valls, H. Bahlouli,
Phys. Rev. B{\bf 45} (1992) 13035.
\bibitem{lsg_rg} I. N\'andori and K. Sailer, to be published in Phil. Mag.,
see also e-print hep-th/0508033.
\bibitem{csmag} I. N\'andori, K. Vad, S. M\'esz\'aros, J. Hakl, B. Sas, 
Czech. J. Phys. {\bf 54} (2004) D481. 
\bibitem{kalman}K. Vad, S. M\'esz\'aros, I. N\'andori, B. Sas, 
to be published in Phil. Mag., 
see also e-print cond-mat/0508146;
K. Vad, S. M\'esz\'aros, B. Sas, to be published in Physica C,
see also e-print cond-mat/0508184.
\bibitem{DeSc1997} D. Delpenich, J. Schechter, 
Int. J. Mod. Phys. A{\bf 12} (1997) 5305.
\bibitem{SmVe1996} A. Smilga, J. J. M. Verbaarschot, 
Phys. Rev. D{\bf 54} (1996) 1087.
\bibitem{wh}F. J. Wegner, A. Houghton, Phys. Rev. A{\bf 8} (1973) 401.
\bibitem{Co1973} S. Coleman, Commun. Math. Phys. {\bf 31} (1973) 259.
\bibitem{Co1975} S. Coleman, Phys. Rev. D{\bf 11} (1975) 2088.
\bibitem{Co1976} S. Coleman, Ann. Phys. {\bf 101} (1976) 239.
\bibitem{sg3}I. N\'andori, K. Sailer, U. D. Jentschura, G. Soff,
Phys. Rev. D{\bf 69} (2004) 025004; J. Phys. G{\bf 28}  (2002) 607.
\end{thebibliography}
\end{document}